\documentclass[12pt]{article}
\usepackage{fullpage}
\usepackage{cite}
\usepackage{amsmath}
\usepackage{amssymb}
\usepackage{epsfig}
\title{}
\date{}
\def\para{\\ [-2mm]}

\def\cAga{ {\cal A}_{\rm gauge} }
\def\cAgv{ {\cal A}_{\rm grav} }
\def\cA{ {\cal A} }
\def\KK{ {\bf K} }
\def\LL{ {\bf L} }
\def\Pf { {\rm Pf~} }

\def\tepsilon{ \tilde{\epsilon}  }
\def\Epsilon{  {\cal E} }
\def\tEpsilon{ \tilde{\cal E}   }

\def\fthree {   \Tr ( T^{\textsf{a}_{1}}T^{\textsf{a}_{2}}  T^{\textsf{a}_{4}} ) }

\def\epseps#1#2{ \epsilon_{#1} \cdot \epsilon_{#2} }
\def\epsk#1#2{ \epsilon_{#1} \cdot k_{#2} }
\def\kk#1#2{ k_{#1} \cdot k_{#2} }

\def\sigmaset{  \{ \sigma \} }
\def\sumset{  \sum_{   \sigmaset \in {\rm solutions}  }   }
\def\bigsubset{  \{ K, \sigma \} }
\def\smallsubset{  \{ k, \sigma \} }
\def\bigset{  \{ \Epsilon, K, \sigma \} }
\def\tbigset{  \{ \tEpsilon, K, \sigma \} }
\def\smallset{  \{ \epsilon, k, \sigma \} }
\def\tsmallset{  \{ \tepsilon, k, \sigma \} }

\def\SL2C{\mathrm{SL}(2,\mathbb{C})}
\def\Tr {\mathop{\rm Tr}\nolimits}

\def\be{\begin{equation}}
\def\ee{\end{equation}}
\def\ba{\begin{eqnarray}}
\def\ea{\end{eqnarray}}
\def\nl{\nonumber\\}

\def\eqn#1{eq.~(\ref{#1})} \def\Eqn#1{Equation~(\ref{#1})}
\def\eqns#1#2{eqs.~(\ref{#1}) and~(\ref{#2})}
\def\Eqns#1#2{Eqs.~(\ref{#1}) and~(\ref{#2})}

\def\id{  {\mathsf{1}\kern -3pt \mathsf{l} } }
\def\half{  {1\over 2} }
\def\ie{{i.e.}}

\begin{document}
\bibliographystyle{nb}

\titlepage
\begin{flushright}
BOW-PH-160\\
\end{flushright}

\vspace{3mm}

\begin{center}
{\Large\bf\sf
CHY representations for gauge theory and gravity amplitudes  \\ [1mm]
with up to three massive particles
}

\vskip 1.5cm

{\sc
Stephen G. Naculich\footnote{
Research supported in part by the National Science
Foundation under Grant No.~PHY14-16123.}
}

\vskip 0.5cm
{\it
Department of Physics\\
Bowdoin College\\
Brunswick, ME 04011, USA
}

\vspace{5mm}
{\tt
naculich@bowdoin.edu
}
\end{center}

\vskip 1.5cm

\begin{abstract}

We show that a wide class of tree-level scattering amplitudes 
involving scalars, gauge bosons, and gravitons, 
up to three of which may be massive,
can be expressed in terms of a Cachazo-He-Yuan representation
as a sum over solutions of the scattering equations.
These amplitudes, 
when expressed in terms of the appropriate kinematic invariants, 
are independent of the masses and therefore identical to
the corresponding massless amplitudes.

\end{abstract}

\vspace*{0.5cm}

\vfil\break
\section{Introduction}
\setcounter{equation}{0}

The new representations for tree-level gauge and gravitational amplitudes
in arbitrary spacetime dimensions
discovered by Cachazo, He, and Yuan \cite{Cachazo:2013hca,Cachazo:2013iea},
while as yet lacking a quantum-field-theoretic foundation,
have illuminated and unified many previous results.
The tree-level $n$-gluon amplitude for pure Yang-Mills theory 
(up to an overall factor proportional to $g_{\rm YM}^{n-2}$)
is given by 
\be
\cAga 
= 
(-1)^{n-1} 
\sumset
\frac{ C(\sigmaset) ~E(\smallset) }{  \det'\Phi (\smallsubset)  } 
\label{chygauge}
\ee
where 
$\sigmaset$ is a set of $n$ points on $\mathbb{C} \mathbb{P}^1$,
$C(\sigmaset)$ depends on group theory structure constants, 
$E(\smallset)$ depends on the momenta and polarizations of the gluons,
and 
$\det'\Phi (\smallsubset)$ is a Jacobian factor defined in sec.~\ref{sec:double}.
The sum is over the $(n-3)!$ solutions of the scattering equations 
\cite{Cachazo:2013gna}
\be
 \sum_{b\neq a} \frac{k_a \cdot k_b}{\sigma_{a}-\sigma_{b}}~=~0,
\qquad \sigma_a \in \mathbb{C} \mathbb{P}^1,
\qquad    a = 1, \cdots, n \,.
\label{scatt}
\ee
The structure of \eqn{chygauge} as a sum 
over products of color- and kinematic-dependent factors 
is closely related to BCJ color-kinematic duality \cite{Bern:2008qj}.
For $n=4$, the scattering equations have a single solution,
so that \eqn{chygauge} implies that the four-gluon amplitude factors 
into a color-dependent term and a polarization-dependent term.
The factorization of a wide class of four-point gauge theory amplitudes 
was first recognized over thirty years ago \cite{Zhu:1980sz,Goebel:1980es}
following the discovery that radiative gauge theory amplitudes
vanish in certain kinematic 
regions \cite{Mikaelian:1977ux,Brown:1979ux,Mikaelian:1979nr,Grose:1980cc,Brodsky:1982sh,Brown:1982xx}.
\para

The tree-level gravity amplitude 
(up to an overall factor proportional to $\kappa^{n-2}$) 
is 
\be
\cAgv 
= 
(-1)^{n-1} 
\sumset
\frac{ E (\smallset) ~E(\tsmallset) }{  \det'\Phi (\smallsubset)  }
\label{chygrav}
\ee
and its structure clearly echoes the double-copy procedure 
of refs.~\cite{Bern:2008qj,Bern:2010ue,Bern:2010yg}.
The Kawai-Lewellen-Tye (KLT) relations of tree-level string theory \cite{Kawai:1985xq}
in the field theory limit \cite{Bern:1998sv,BjerrumBohr:2010hn}
also follow from \eqn{chygrav}.
In particular, \eqn{chygrav} implies that the four-graviton amplitude 
factors into a pair of gauge-theory partial amplitudes.
This has long been known from the field-theory limit  
of the four-point closed string amplitude \cite{Green:1982sw}
as well as from an explicit field theory calculation \cite{Sannan:1986tz}.
Many other four-point amplitudes involving gravitons 
have also been shown to factor into a pair of gauge theory amplitudes in an arbitrary number 
of spacetime dimensions \cite{Choi:1994ax,Holstein:2006bh,Holstein:2006ry,Bjerrum-Bohr:2014lea}.
\para

In refs.~\cite{Cachazo:2014nsa,Cachazo:2014xea},
Cachazo, He, and Yuan extended their results
to amplitudes for massless particles in a variety of theories
using dimensional reduction and other techniques.
Other work inspired by these developments includes refs.~\cite{Litsey:2013jfa,
Adamo:2013tca, Monteiro:2013rya, Mason:2013sva, Chiodaroli:2013upa,
Berkovits:2013xba, Dolan:2013isa, Adamo:2013tsa, Gomez:2013wza, Kalousios:2013eca,
Stieberger:2014hba, Yuan:2014gva, Weinzierl:2014vwa, Dolan:2014ega, 
Bjerrum-Bohr:2014qwa,He:2014wua, Kol:2014yua, Kol:2014zca, Geyer:2014fka,
Naculich:2014rta,
Schwab:2014xua,
Afkhami-Jeddi:2014fia,Zlotnikov:2014sva,Kalousios:2014uva,
Naculich:2014naa,
Adamo:2014wea,
Lam:2014tga,
Weinzierl:2014ava}.
\para

Most of the attention so far has been focused on amplitudes for massless particles.
The factorization of four-point gauge theory amplitudes 
found in ref.~\cite{Zhu:1980sz,Goebel:1980es}, however,
required that only one of the gauge bosons involved be massless,
and the four-point gravitational amplitudes considered in 
refs.~\cite{Choi:1994ax,Holstein:2006bh,Holstein:2006ry,Bjerrum-Bohr:2014lea}
involved massive particles as well.
This suggests that CHY representations may 
be possible for various amplitudes containing massive particles.
Dolan and Goddard proposed a modification of the scattering equations 
for massive particles with equal mass 
and derived a CHY representation for $\phi^3$ theory \cite{Dolan:2013isa,Dolan:2014ega}.
In ref.~\cite{Naculich:2014naa},
the scattering equations were generalized to arbitrary masses,
and CHY representations were proposed for gauge theory amplitudes 
involving a pair of massive scalar particles in the fundamental representation
and an arbitrary number of gluons, 
as well as for gravitational amplitudes
involving a pair of massive scalar particles 
and an arbitrary number of gravitons.
\para

In this paper, we derive CHY representations for a large class 
of tree-level gauge theory and gravitational amplitudes, 
with up to three massive particles.
We require only that the remaining (massless) particles be flavor-preserving
(as is generally the case).
Furthermore, we show that these amplitudes,
when written in terms of the appropriate kinematic invariants,
are independent of the masses of the particles,
and thus identical to purely massless amplitudes.
Our approach is first to show that all the diagrams contributing to an amplitude 
with up to three massive particles (with the remaining particles being 
flavor-preserving) can be kinematically regarded 
as the dimensional reduction of diagrams of purely massless amplitudes,
with non-zero momenta in the extra dimensions.
Then, by different choices of the polarization vectors,
we can use \eqns{chygauge}{chygrav} in higher dimensions to derive 
CHY representations for mixed amplitudes in lower dimensions,
as was done in ref.~\cite{Cachazo:2014xea}.
In this way, we derive a variety of gauge theory amplitudes involving massive gauge
bosons and scalar particles, and gravitational amplitudes involving
gravitons, massive gauge particles, and massive scalars.
We recover the results of ref.~\cite{Naculich:2014naa} as a special case.
\para

This paper is structured as follows.
In sec.~\ref{sec:propagator}, we show that the propagator matrix
for an amplitude with up to three massive particles has the same rank 
as that for a purely massless amplitude, motivating the search
for a CHY-type representation for such amplitudes.
In sec.~\ref{sec:scattering}, we derive the scattering equations 
for amplitudes with up to three massive particles
from massless scattering equations in higher dimensions. 
In sec.~\ref{sec:double},
we derive a CHY representation for the massive propagator matrix. 
In sec.~\ref{sec:gauge}, 
we dimensionally reduce massless gauge boson amplitudes 
to obtain CHY representations for amplitudes with massive and massless gauge bosons,
and mixed amplitudes containing gauge bosons and adjoint or fundamental scalars.
In sec.~\ref{sec:gravity}, we dimensionally reduce 
pure graviton amplitudes to obtain CHY representations for
mixed amplitudes containing gravitons, massive or massless gauge bosons, 
and massive scalars.
Sec.~\ref{sec:concl} contains our conclusions. 
\para

\section{Propagator matrix}
\setcounter{equation}{0}
\label{sec:propagator}

In this section,
we briefly review the propagator matrix that appears in
the amplitude for massless gauge bosons \cite{Vaman:2010ez}.
We then show that the propagator matrix 
for an $n$-point amplitude with up to three massive particles
(assuming that the remaining $n-3$ particles are flavor-preserving)
is identical to that for an amplitude with all massless particles.
Like the latter,  therefore, it has rank $(n-3)!$. 
This in turn suggests the possibility of finding a CHY representation 
in terms of a sum over the $(n-3)!$ solutions of the scattering equations
for these amplitudes.
\para

We begin with the tree-level $n$-gluon amplitude 
written as a sum over cubic diagrams \cite{Bern:2008qj}
\be
\cAga ~=~ \sum_i   { c_i n_i \over d_i }
\label{cubicdecomposition}
\ee
where $c_i$, $n_i$, and $d_i$ are the color factors,
kinematic numerators, and denominators (arising from the propagators)
associated with the diagram.
The color factors $c_i$ of each of the cubic diagrams appearing in \eqn{cubicdecomposition}
can be rewritten, using Jacobi identities,  in terms of a basis of $(n-2)!$ color factors
\cite{DelDuca:1999ha,DelDuca:1999rs}
\be
c_i ~=~ \sum_{\gamma \in S_{n-2}}  M_{i, 1\gamma n} {\bf c}_{1 \gamma n } 
\label{cMc}
\ee
where $ \gamma $ denotes a permutation of $\{2, \cdots, n-1\}$ and 
\be
{\bf c}_{1 \gamma n } ~\equiv~
{\bf c}_{1 \gamma(2) \cdots \gamma(n-1) n } 
~\equiv~  \sum_{\textsf{b}_1,\ldots,\textsf{b}_{n{-}3}} 
f^{\textsf{a}_{1} \textsf{a}_{\gamma(2)} \textsf{b}_1}
f^{\textsf{b}_{1} \textsf{a}_{\gamma(3)} \textsf{b}_2}\cdots 
f^{\textsf{b}_{n{-}3} \textsf{a}_{\gamma(n{-}1)} \textsf{a}_{n}}  \,.
\label{defc}
\ee
The color factors
${\bf c}_{1 \gamma n }$
are associated with half-ladder diagrams, \ie, diagrams in which there is a 
line connecting particles 1 and $n$ to which each of the other particles
is directly connected.
The assumption of color-kinematic duality \cite{Bern:2008qj} is that the kinematic 
numerators $n_i$ obey the same Jacobi identities as $c_i$, and therefore
can be written in terms of $(n-2)!$ half-ladder numerators
\be
n_i ~=~ \sum_{\gamma \in S_{n-2}}  M_{i, 1\gamma n} {\bf n}_{1 \gamma n }   \,.
\label{nMn}
\ee
\Eqns{cMc}{nMn} can be used to rewrite the amplitude (\ref{cubicdecomposition}) as
\be
\cAga
~=~  \sum_{\gamma \in S_{n-2}}  \sum_{\delta \in S_{n-2}}  
{\bf n}_{1 \gamma n } ~m( 1 \gamma n| 1 \delta n) ~{\bf c}_{1 \delta n }  
\label{nmc}
\ee
where we define the entries of the 
$(n-2)! \times (n-2)!$ propagator matrix \cite{Vaman:2010ez}
\be 
m( 1 \gamma n| 1 \delta n) 
~=~ 
\sum_i 
{M_{i,1\gamma n} M_{i, 1\delta n} \over d_i} 
\label{propagatormatrix}
\ee
as the sum (weighted by the denominators $1/d_i$ of the diagrams) 
over those cubic diagrams that contribute to both 
${\bf c}_{1 \gamma n }  $ and ${\bf c}_{1 \delta n }  $.
\para

\begin{figure}
\centerline{\epsfxsize=10.0cm \epsffile{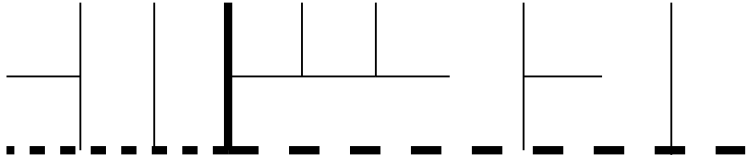}}
\caption{A cubic diagram in which thin lines denote massless propagators, 
while the thick lines denote massive propagators (with possibly different masses)}
\end{figure}

More generally, we can define the propagator matrix for a gauge theory amplitude
with particles other than gluons by using group theory relations 
to rewrite the color factors in terms of a half-ladder basis
(for example, see ref.~\cite{Naculich:2014naa} for amplitudes with
a pair of particles in the fundamental representation).
Now consider an $n$-point amplitude with up to three massive particles,
with masses $m_1$, $m_2$, and $m_n$,
where the remaining $(n-3)$ particles are flavor-preserving.
Each of the individual cubic diagrams
contributing to the sum (\ref{cubicdecomposition})
can be viewed as a skeleton 
with three massive lines meeting at a vertex
fleshed out with branches of massless lines emanating from each 
of the massive lines and from other massless lines; an example is shown in Fig.~1.
The denominator factors associated with the side branches
are of the form 
\be 
\left( \sum_{a \subset S}  k_a\right)^2
~=~ \sum_{ {a,b \subset S \atop a\neq b}}  2 k_a \cdot k_b
\label{masslessinverse}
\ee
where $S$ is some subset of the (massless)  momenta $\{ 3, \cdots, n-1 \}$.
The denominator factors associated with the three massive lines are 
of the form
\be
 \left(k_c + \sum_{a \subset S}  k_a\right)^2 - m_c^2
~=~ \left(\sum_{a \subset S}  k_a\right)^2  + \sum_{a \subset S} 2 k_c \cdot k_a  \,,
\qquad    c \in \{1, 2, n \} \,.
\label{massiveinverse}
\ee 
Thus, all of the inverse 
propagators (\ref{masslessinverse}) and (\ref{massiveinverse}) 
are given by linear combinations of the 
kinematic invariants
\be
k_1 \cdot k_a, \qquad k_2 \cdot k_a,  \qquad
k_n \cdot k_a, \qquad k_a \cdot k_b,\qquad\qquad
a, b \in \{3, \cdots, n-1\}
\label{invariants}
\ee
with no explicit dependence on the masses
$m_1$, $m_2$, and $m_n$.
(Of course, momentum conservation can be used to eliminate $k_n \cdot k_a$,
leaving $n(n-3)/2$ independent invariants.)
\para

Because the inverse propagators $d_i$,
when written in terms of the invariants (\ref{invariants}),
do not depend on the masses, 
neither does the propagator matrix (\ref{propagatormatrix}).
And since the $(n-2)! \times (n-2)!$ propagator matrix 
for amplitudes with massless particles has rank $(n-3)!$ as a consequence
of momentum conservation \cite{Vaman:2010ez},
the same is true for the propagator matrix
for amplitudes with up to three massive particles.
This suggests the possibility of expressing it as a sum over the 
$(n-3)!$ solutions of the scattering equations, which we will do
in the next two sections.
\para

On the other hand, 
if there are four or more massive particles in the amplitude,
the rank of the propagator matrix will generally be larger than $(n-3)!$.
For example, the amplitude for four massive particles, 
one of which is flavor-preserving,
has a $2 \times 2$ propagator matrix with non-vanishing determinant,
and thus rank two.
\para

\section{Scattering equations}
\setcounter{equation}{0}
\label{sec:scattering}

In order to obtain CHY representations for amplitudes with massive particles,
we first need the appropriate scattering equations.
In this section we show that the scattering equations 
for an $n$-point amplitude with up to three massive particles are identical to, 
and therefore have the same set of solutions as,
the scattering equations for massless particles.
We also show that the massive scattering equations for amplitudes
with up to three massive particles
can be derived from the massless scattering equations in higher dimensions.
\para

The scattering equations for massless particles are given in 
\eqn{scatt}. 
In refs.~\cite{Dolan:2013isa,Dolan:2014ega},
Dolan and Goddard proposed a modification 
of these equations for massive particles with equal mass.
The scattering equations were subsequently generalized \cite{Naculich:2014naa}
for particles with distinct masses $ m_a $ :
\be
f_a \equiv \sum_{b\neq a} \frac{k_a \cdot k_b + \Delta_{ab}} {\sigma_{a}-\sigma_{b}}~=0, 
\qquad \sigma_a \in \mathbb{C} \mathbb{P}^1,
\qquad    a = 1, \cdots, n \,.
\label{massivescatt}
\ee
This set of equations is invariant under $\SL2C$  transformations
\be
\sigma \longrightarrow {A \sigma + B \over C \sigma + D}, \qquad A D-B C = 1
\label{sl2c}
\ee
provided $\sum_{a=1}^n k_a =0$ and that $\Delta_{ab}$ satisfy the constraints
\be
\Delta_{ab} = \Delta_{ba}, \qquad \sum_{b\neq a} \Delta_{ab} = m_a^2 \,.
\label{deltaconstraint} 
\ee
Furthermore,  when the constraints (\ref{deltaconstraint}) are satisfied,
the three sums
\be
\sum_{a=1}^n f_a, \qquad 
\sum_{a=1}^n \sigma_a f_a, \qquad
\sum_{a=1}^n \sigma_a^2 f_a
\ee
vanish identically (\ie, before imposing $f_a=0$),
which implies that only $n-3$ of the massive scattering equations (\ref{massivescatt})
are independent.
\para

Now consider an amplitude with three massive particles, $m_1$, $m_2$, and $m_n$.
The constraints (\ref{deltaconstraint}) can be satisfied by 
choosing
\ba
\Delta_{12} ~=~ \Delta_{21} &= \half \left( m_1^2 + m_2^2 - m_n^2\right)\,, \nl
\Delta_{2n} ~=~ \Delta_{n2} &= \half \left( m_2^2 + m_n^2 - m_1^2\right)\,, \nl
\Delta_{n1} ~=~ \Delta_{1n} &= \half \left( m_n^2 + m_1^2 - m_2^2\right) 
\label{threeDeltas}
\ea
with the remaining $\Delta_{ab}$ vanishing.
Since only $n-3$ of the $n$ scattering equations (\ref{massivescatt}) are independent,
we can eliminate those for $a=1$, 2, and $n$. 
Since $\Delta_{ab}$ vanishes in the remaining equations (for $a= 3, \cdots, n-1$),
they are identical to the scattering equations for massless particles,
and hence have the same set of $(n-3)!$ solutions,
when expressed in terms of the kinematic invariants  (\ref{invariants}).
This will be crucial in the remainder of this paper.
\para

Finally, we show that the massive scattering equations (\ref{massivescatt})
are precisely the equations that arise if one regards the massive particles 
as massless particles in a higher-dimensional space.
In particular, the momentum $K_a$ for a massless particle in $(d+M)$ dimensions
can be written 
\be
K_a ~=~ (k_a | \kappa_a)
\ee
where $k_a$ and $\kappa_a$ are the components of momentum in $d$- and $M$-dimensional
subspaces respectively.\footnote{We will use a mostly-minus metric for $K_a$ and $k_a$, 
but an all-plus metric for the internal components $\kappa_a$.} 
We will refer to $\kappa_a$ as the {\it internal momentum}
of a particle;
the mass of the particle in $d$ dimensions is 
given by $m_a^2 = \kappa_a^2$.
The massless scattering equations in $(d+M)$ dimensions 
\be
 \sum_{b\neq a} \frac{K_a \cdot K_b}{\sigma_{a}-\sigma_{b}}~=~0,
\qquad    a = 1, \cdots, n
\label{Kscatt}
\ee
imply that the scattering equations (\ref{massivescatt}) hold in $d$ dimensions with
\be
\Delta_{ab} ~=~   -\, \kappa_a \cdot \kappa_b
\label{Deltakappa}
\ee
and the constraints (\ref{deltaconstraint}) hold automatically by virtue of momentum conservation in the internal dimensions.
\para

If $M=1$, then momentum conservation in the internal space requires
$\sum_{a=1}^n  (\pm m_a) = 0$ (for some choice of signs),
which is too restrictive for our purposes.
If $M\ge 2$, however, there is 
no constraint on the masses of the particles.
\para

It must be stressed, however, that by no means can every massive amplitude 
be obtained from the dimensional reduction of a massless amplitude
because momentum conservation (including conservation of internal momentum)
must hold at each vertex of a diagram, 
placing severe constraints on the masses of the intermediate particles 
in any given diagram.  
A simple example will suffice to explain  the difficulty.
Consider the four-point amplitude involving $\bar{u}$, $d$, $W^+$, and $Z^0$.
Since the $Z^0$ is flavor- and therefore mass-preserving, 
conservation of momentum at any vertex involving the $Z^0$
implies that $(\kappa_i + \kappa_Z)^2 = \kappa_i^2$, 
where $\kappa_i$ is the internal momentum for $\bar{u}$, $d$, or $W^+$.
Together with overall internal momentum conservation ($\kappa_Z = -\sum_i \kappa_i$),
this implies that $\kappa_Z^2 = 0$, which implies that the $Z_0$ must be massless.
\para

For the amplitudes considered in this paper, however, 
with (up to) three massive particles, and the remaining (massless) particles
flavor-preserving,
internal momentum conservation at each vertex involving a massless 
($\kappa=0$) flavor-conserving particle holds automatically,
and internal momentum conservation at the vertex involving the three massive particles,
$\kappa_1 + \kappa_2 + \kappa_n = 0$, is easily satisfied for appropriate
choices of $\kappa_i$.
\para

\section{Double-partial amplitudes}
\setcounter{equation}{0}
\label{sec:double}

In this section, we show that the propagator matrix (\ref{propagatormatrix}) 
for an amplitude with up to three massive particles 
can be written as a sum over the solutions of the scattering equations (\ref{massivescatt}).
\para

First we recall from sec.~{\ref{sec:scattering}} that 
kinematically we can regard the scattering of $n$ particles 
(up to three of them massive and the rest flavor-preserving) in $d$ dimensions
as the scattering of $n$ massless particles in $d+M$ dimensions,
with the same propagator matrix.
\para

Next, recall that  Cachazo, He, and Yuan \cite{Cachazo:2013iea}
demonstrated that the entries of the propagator matrix 
for massless amplitudes can be interpreted as 
double-partial amplitudes of a theory of massless scalar particles 
transforming in the adjoint of the color group $U(N) \times U(\tilde{N})$
and presented a compact formula for computing them.
First they defined an $n \times n$ matrix $\Phi (\bigsubset) $ with entries 
\be
\Phi_{ab} ~=~ 
\frac{2 K_a \cdot K_b }{(\sigma_a-\sigma_b)^2}, \quad  a\neq b;
\qquad\qquad 
\Phi_{aa} ~=~ 
 -\sum_{c\neq a}\frac{2 K_a \cdot K_c }{(\sigma_a-\sigma_c)^2} 
\ee
where $K_a$ are the $(d+M)$-dimensional momenta of massless particles.
When the scattering equations (\ref{Kscatt}) are satisfied, this matrix satisfies 
$
\sum_{a=1}^n \Phi_{ab} 
= \sum_{a=1}^n \sigma_a \Phi_{ab} 
= \sum_{a=1}^n \sigma_a^2 \Phi_{ab} = 0$,
and therefore has rank $(n-3)$.
Define the nonsingular matrix $\Phi^{ijk}_{pqr}$ by removing rows 
$i$, $j$, and $k$, and columns $p$, $q$, and $r$,
and let $|\Phi|^{ijk}_{pqr}$ be its signed determinant.
The combination
\be
{\det}'\Phi (\bigsubset) \equiv  \frac{|\Phi|^{ijk}_{pqr}}{(\sigma_{p,q}\sigma_{q,r}\sigma_{r,p})(\sigma_{i,j}\sigma_{j,k}\sigma_{k,i})}
\label{detPhi}
\ee
where $\sigma_{a,b} \equiv \sigma_a-\sigma_b$,
was then shown to be independent of the choices of removed rows and columns.
CHY then demonstrated that the entries of the propagator matrix can be computed using
\be
m( 1 \gamma n | 1 \delta n) 
~=~ 
\sumset 
\frac{(-1)^{n-1} }{\det'\Phi (\bigsubset) }
\frac{1}
{
(\sigma_{1,\gamma(2)}\cdots\sigma_{\gamma(n-1),n}\sigma_{n,1})
(\sigma_{1, \delta(2)}\cdots\sigma_{ \delta(n-1),n}\sigma_{n,1})
}
\label{doublepartialK}
\ee
where the sum is over the $(n-3)!$ solutions of the scattering equations (\ref{Kscatt}).
\para

Now we need to rewrite \eqn{doublepartialK} in terms of $d$-dimensional quantities.
Since $ \det'\Phi$ is independent of the choices of removed rows and columns,
we can choose to remove the rows and columns associated with the massive particles
(1, 2, and $n$).   
Since the remaining rows and columns correspond to massless particles (with $\kappa_a=0$),
the numerators in $\Phi_{12n}^{12n}$ are
given by $K_a \cdot K_c = k_a \cdot k_c $
with $a \in  \{3, \cdots, n-3\}$ and $c \in  \{1, \cdots, n  \}$,
which are precisely the set of kinematic invariants (\ref{invariants}).
This implies that\footnote{Clearly
this would not be the case if four or more of the particles were massive.}
\be
{\det}'\Phi ( \bigsubset ) ~=~ 
{\det}'\Phi ( \smallsubset )  \,.
\ee
provided that the determinant is expressed in terms of the invariants (\ref{invariants}).
Furthermore, as we saw in sec.~\ref{sec:scattering}, 
the massless scattering equations in $d+M$ dimensions
(\ref{Kscatt}) are precisely equivalent to the massive scattering equations 
in $d$ dimensions (\ref{massivescatt}),
and therefore have the same solutions.
Thus, the propagator matrix for the massive amplitude is given by 
\be
m( 1 \gamma n | 1 \delta n) 
~=~ 
\sumset 
\frac{(-1)^{n-1} }{\det'\Phi (\smallsubset) }
\frac{1}
{
(\sigma_{1,\gamma(2)}\cdots\sigma_{\gamma(n-1),n}\sigma_{n,1})
(\sigma_{1, \delta(2)}\cdots\sigma_{ \delta(n-1),n}\sigma_{n,1})
}
\label{doublepartialk}
\ee
where the sum is over the solutions of the scattering equations (\ref{massivescatt}).
\para

Finally, the expression (\ref{doublepartialk}) has no explicit dependence on the masses, 
nor does it have any implicit dependence on the masses
since the solutions $\{ \sigma \}$ 
of the scattering equations,
when expressed in terms of the kinematic invariants (\ref{invariants}), 
are independent of the masses.
Hence, \eqn{doublepartialk} for an amplitude with up to three massive particles
is the same as that for massless amplitudes, 
as we already saw in sec.~\ref{sec:propagator}.
\para

\section{Gauge theory amplitudes}
\setcounter{equation}{0}
\label{sec:gauge}

In this section, 
we demonstrate that various gauge theory amplitudes involving 
up to three massive particles can be expressed, like those for 
massless particles,  as a sum over the $(n-3)!$ solutions of the 
scattering equations.
Moreover, the expressions for these amplitudes, 
when written in terms of the kinematic invariants (\ref{invariants}),
are independent of the masses of the particles. 
\para

We begin in $d+M$ dimensions with the CHY representation
for the tree-level scattering amplitude of 
$n$ massless gauge bosons \cite{Cachazo:2013hca,Cachazo:2013iea}
\be
\cAga ~=~ 
(-1)^{n-1} 
\sumset
\frac{ C(\sigmaset) ~E(\bigset) }{  \det'\Phi (\bigsubset)  } \,.
\label{sumCE}
\ee
Here $\Epsilon_a$ and $K_a$  are the $(d+M)$-dimensional polarizations and 
momenta of the gauge bosons,
and the sum is over the solutions of the scattering equations (\ref{Kscatt}).
The first factor in the numerator is 
\be
C(\sigmaset) ~=~ 
\sum_{\gamma \in S_{n-2} }
\frac{  {\bf c}_{1 \gamma n} }
{\sigma_{1,\gamma(2)}\cdots\sigma_{\gamma(n-1),n}\sigma_{n,1}}
\label{defC}
\ee
where
${\bf c}_{1 \gamma n} $ was defined in \eqn{defc},
and $\det' \Phi$ was defined in \eqn{detPhi}. 
It is apparent that color-kinematic dual numerators ${\bf n}_{1 \gamma n} $ 
can be used to construct 
\be
E(\bigset) ~=~ 
\sum_{\delta \in S_{n-2} }
\frac{  {\bf n}_{1 \delta n} }
{\sigma_{1,\delta(2)}\cdots\sigma_{\delta(n-1),n}\sigma_{n,1}} 
\label{Enum}
\ee
so that \eqn{doublepartialK} implies that \eqns{sumCE}{nmc} are equivalent.
The novelty of the CHY approach, however, is in providing an alternative definition 
for $E(\bigset)$ in terms of the Pfaffian of a
matrix, so that \eqn{sumCE} yields the gauge theory amplitudes directly
without resort to Feynman diagrams.  \para

One first defines a $2n \times 2n$ matrix 
\be
\Psi (\bigset) ~=~ \left(
         \begin{array}{cc}
           A &  -C^{\rm T} \\
           C & B \\
         \end{array}
       \right)
\label{defPsi}
\ee
where $A$, $B$, and $C$ are the $n \times n$ submatrices 
\be
A_{ab} = 
\begin{cases} 
\displaystyle \frac{K_a \cdot K_b }{\sigma_{a}-\sigma_{b}}, & a\neq b;\\[5mm]
\displaystyle \quad ~~ 0, & a=b;
\end{cases} 
~~~~
B_{ab} = 
\begin{cases} 
\displaystyle \frac{ \Epsilon_a\cdot \Epsilon_b}{\sigma_{a}-\sigma_{b}}, & a\neq b;\\[5mm]
\displaystyle \quad ~~ 0, & a=b;
\end{cases}
~~~~
C_{ab} = 
\begin{cases} 
\displaystyle \frac{ \Epsilon_a\cdot K_b}{\sigma_{a}-\sigma_{b}}, &a\neq b;\\[5mm]
\displaystyle -\sum_{c\neq a}\frac{ \Epsilon_a\cdot K_c}{\sigma_{a}-\sigma_{c}}, &a=b.
\end{cases}
\label{submatrices}
\ee
The matrix $\Psi$ has two null vectors, 
so to obtain a nonvanishing Pfaffian,
it is necessary to remove two of the first $n$ rows and columns.
It was shown in ref.~\cite{Cachazo:2013hca} that the resulting 
``reduced   Pfaffian''
is independent 
(when the scattering equations are satisfied)
of the choice of rows and columns deleted,
but for our purposes (in order to make manifest the mass independence of the result)
it will prove advantageous to choose ones
corresponding to two of the massive particles.
We therefore define $\Psi_{1,n}$ as the matrix obtained by deleting
the 1st and $n$th rows and columns from $\Psi$,
in which case the CHY prescription is
\be
E(\bigset) ~=~ \frac{(-1)^{n+1} } {\sigma_{1,n} } \Pf  \Psi_{1,n} (\bigset) \,.
\label{EPsiK}
\ee
It was shown in refs.~\cite{Cachazo:2013hca,Dolan:2013isa} that, with this definition,
\eqn{sumCE} computes the amplitudes for massless gauge bosons 
in $(d+M)$ dimensions.
\para

\medskip\noindent{{\bf Amplitudes for massive gauge bosons}}\smallskip

We now use dimensional reduction
to obtain the amplitude for massless and 
(up to three) massive gauge bosons in $d$ dimensions.
As in sec.~\ref{sec:scattering}, we set $K_a = (k_a | \kappa_a)$,
where the massless bosons ($a=3, \cdots, n-1$) have $\kappa_a=0$.
We restrict all of the $(d+M)$-dimensional 
polarization vectors $\Epsilon_a$ to the $d$-dimensional subspace:
$\Epsilon_a = (\epsilon_a |  0, 0, \cdots )$.
All of the $(d+M)$-dimensional dot products 
in the matrices $B$ and $C$ reduce to $d$-dimensional dot products,
$\Epsilon_a \cdot \Epsilon_b =\epsilon_a \cdot \epsilon_b$
and $\Epsilon_a \cdot K_b =\epsilon_a \cdot k_b$,
whereas those in $A$
become $K_a \cdot K_b = k_a \cdot k_b -\kappa_a\cdot \kappa_b$.
Every matrix element $A_{ab}$ that appears in $\Psi_{1,n}$
(from which the 1st and $n$th rows and columns of $\Psi$ have been eliminated)
has either $\kappa_a=0$ or $\kappa_b=0$,
so $K_a \cdot K_b = k_a \cdot k_b$,
which belongs to the set of kinematic invariants (\ref{invariants}).
Thus,  \eqn{EPsiK} becomes
\be
E(\bigset) ~=~ \frac{(-1)^{n+1} } {\sigma_{1,n} } \Pf  \Psi_{1,n} (\smallset) \,.
\label{EPsik}
\ee
so the $d$-dimensional amplitude for $n$ gauge bosons, up to three of which may be 
massive, is 
\be
\cA 
~=~ 
\sumset 
{ C(\sigmaset) \over \det'\Phi (\smallsubset)   }  {\Pf  \Psi_{1,n} (\smallset) \over   \sigma_{1,n}} \,.
\label{massivegauge}
\ee
where  $\sigmaset$ are solutions of the scattering equations (\ref{massivescatt}).
We showed in sec.~\ref{sec:scattering} that these solutions,
when expressed in terms of the invariants (\ref{invariants}),
are independent of the masses,
and in sec.~\ref{sec:double} that $\det' \Phi$
is also independent of the masses.
Hence, 
we conclude that the scattering amplitude (\ref{massivegauge}) 
for $n$ gauge bosons, up to three of which may be massive, 
is identical to that for $n$ massless gauge bosons,
provided it is expressed in terms of the invariants (\ref{invariants}).
\para

Let us illustrate this explicitly for the four-gauge-boson amplitude, with $m_3=0$. 
We use $\SL2C$ invariance to set
$\sigma_1=0$, $\sigma_2=1$, and $\sigma_4 \to \infty$,
then evaluate the various components of \eqn{massivegauge}
on the single solution 
$\sigma_3 = -k_1 \cdot k_3/k_3 \cdot k_4$
of the $n=4$ scattering equations.
Taking $\sigma_4$ large but finite, we obtain
\ba
{\det}' \Phi (\smallsubset) &\longrightarrow&  {2 (\kk{3}{4} )^3 \over \sigma_4^4~\kk{1}{3} ~\kk{2}{3} } 
\label{Phifour}  \\
C(\sigmaset) & \longrightarrow&  - ~ {   (\kk{3}{4} )^2 \over \sigma_4^2 ~\kk{2}{3} } 
\left[ { {\bf c}_{1234} \over \kk{3}{4} } +  { {\bf c}_{1324} \over \kk{1}{3} } \right]
\label{Cfour} \\
\Pf  \Psi_{1,4} (\smallset) 
& \longrightarrow &  {   \kk{3}{4}\over \sigma_4 ~\kk{1}{3} ~\kk{2}{3}}  \KK
\label{Efourgauge}
\ea
where 
${\bf c}_{1234} = f^{\textsf{a}_{1} \textsf{a}_{2} \textsf{b}} f^{\textsf{b} \textsf{a}_{3} \textsf{a}_4}$
and
${\bf c}_{1324} = f^{\textsf{a}_{1} \textsf{a}_{3} \textsf{b}} f^{\textsf{b} \textsf{a}_{2} \textsf{a}_4}$,
with $f^{\textsf{a} \textsf{b} \textsf{c}}$ the three-gauge-boson vertex, 
and $\KK$ is the totally permutation-symmetric expression that 
appears in the scattering amplitude for four open strings 
(cf. eqn.~(7.4.42) of ref.~\cite{Green:1987sp})
\ba
\KK &=& 
- \Big[ 
     (\kk{1}{ 3} ~\kk{2}{ 3} ) ~\epseps{1}{ 2} ~\epseps{3}{ 4} \nl
&& + ( \kk{2}{ 3} ~\kk{3}{ 4}) ~\epseps{1}{ 3} ~\epseps{2}{ 4} \nl
&&+ ( \kk{1}{ 3} ~\kk{3}{ 4} ) ~\epseps{1}{ 4} ~\epseps{2}{ 3} \nl
&& +(\kk{1}{ 3} ~\epsk{1}{ 4}~ \epsk{2}{ 3} +\kk{2}{ 3} ~\epsk{1}{ 3} ~\epsk{2}{ 4}  )~\epseps{3}{ 4}  \nl
&& +(\kk{2}{ 3}~ \epsk{1}{ 2} ~\epsk{3}{ 4} + \kk{3}{ 4} ~\epsk{1}{ 4} ~\epsk{3}{ 2} )~\epseps{2}{ 4}  \nl
&& +(\kk{1}{ 3} ~ \epsk{1}{ 2}~ \epsk{4}{ 3} + \kk{3}{ 4}~ \epsk{1}{ 3}~ \epsk{4}{ 2})~\epseps{2}{ 3}  \nl
&& +( \kk{1}{ 3}~ \epsk{2}{ 1}~ \epsk{3}{ 4} +\kk{3}{ 4}~ \epsk{2}{ 4} ~\epsk{3}{ 1} )~\epseps{1}{ 4}   \nl
&& +( \kk{2}{ 3}~ \epsk{2}{ 1}~ \epsk{4}{ 3}+ \kk{3}{ 4}~ \epsk{2}{ 3} ~\epsk{4}{ 1} ) ~\epseps{1}{ 3}  \nl
&&+( \kk{1}{ 3}~ \epsk{3}{ 2}~ \epsk{4}{ 1} + \kk{2}{ 3}~ \epsk{3}{ 1}~ \epsk{4}{ 2})~\epseps{1}{ 2}   
\Big] \,.
\label{defK}
\ea
Assembling the pieces, and taking the limit $\sigma_4 \to \infty$, we  obtain 
\be
\cA_{gggg} ~=~ { 1 \over 2 \kk{2}{3} } \left[ { {\bf c}_{1234} \over \kk{3}{4} } +  { {\bf c}_{1324} \over \kk{1}{3} } \right] \KK, \qquad\qquad \hbox{ when } m_3=0 \,.
 \label{fourgauge}
\ee
This expression exhibits the expected factorization 
into color-dependent and polarization-dependent factors
\cite{Zhu:1980sz,Goebel:1980es}.
\para

Consider two amplitudes involving massive gauge bosons: 
$ {W \gamma \to W\gamma }$ and ${W \gamma \to W Z^0}$.
In each case we have ${\bf c}_{1234} = {\bf c}_{1324}$,  so that
\eqn{fourgauge} reduces to 
\be
{\cal A}_{W \gamma \gamma W} ~\propto~ 
 { e^2 \over \kk{1}{3} ~ \kk{3}{4}  }\KK,
\qquad\qquad
{\cal A}_{W Z^0 \gamma W} ~\propto~ 
 { e^2 \cot \theta_W \over \kk{1}{3} ~ \kk{3}{4}  }\KK
\label{Wcompton}
\ee
both of which can be verified by a Feynman diagram calculation.\footnote{
Holstein \cite{Holstein:2006ry}   
calculated the Compton scattering amplitude from a massive spin-one 
particle with an arbitrary value $g$ for its magnetic moment.   
Our result (\ref{Wcompton}) agrees with his when $g=2$, 
the standard model magnetic moment for the $W$ boson.}
A Feynman diagram calculation also shows that \eqn{fourgauge}
is {\it not} valid for $ W Z^0 \to W Z^0$,
which is not surprising since this amplitude violates the condition $m_3 =0$. 
\para

\medskip\noindent{{\bf Amplitudes for two adjoint scalars and $n-2$ gauge bosons}}\smallskip

Next we turn to the amplitude 
for two adjoint scalars (with masses $m_1$ and $m_n$) and 
$n-2$ gauge bosons, one of which ($m_2)$ may be massive.
This amplitude may be obtained by dimensional reduction of \eqn{sumCE} 
with $\Epsilon_1 = \Epsilon_n = (0 | 1,0,\cdots )$
and $\Epsilon_a = (\epsilon_a | 0, 0, \cdots)$ for $a=2, \cdots, n-1$,
as was done in ref.~\cite{Cachazo:2014xea}.
In that case, where all the particles were massless,
$\Psi$ became block diagonal,  
but the situation here is slightly more complicated
because 
$K_a = (k_a | \kappa_a)$
with $\kappa_a \neq 0$ for $a=1$, 2, and $n$.
\para

We can evaluate the Pfaffian of a $2m \times 2m$ matrix $Z$ by using the relation 
\be
\Pf Z ~=~ \sum_{p=1}^{2m} (-1)^{p+1} z_{p,2m} \Pf Z_{p,2m} 
\label{PfZ}
\ee
where $z_{pq}$ is a matrix element of $Z$,
and we define $Z_{pq}$ as the matrix obtained from $Z$ by omitting 
the $p$th and $q$th rows and columns. 
Observe that only four entries in the $(2n)$th column of $\Psi$ are
non-zero: $\psi_{1,2n}$, $\psi_{2,2n}$, $\psi_{n,2n}$, and $\psi_{n+1,2n}$.
Two of these, $\psi_{1,2n}$ and $\psi_{n,2n}$, have already been 
omitted from $\Psi_{1,n}$,
leaving only two terms in the sum (\ref{PfZ}) for $Z=\Psi_{1,n}(\bigset)$.
Next observe that only four entries in the $(n+1)$th column of $\Psi$ are
non-zero: $\psi_{1,n+1}$, $\psi_{2,n+1}$, $\psi_{n,n+1}$, and $\psi_{2n,n+1}$.
All four of these have been omitted from ${\Psi}_{1,2,n,2n}$, 
and the Pfaffian of a matrix with an entire column of zeros vanishes.
Thus, only one term in the sum (\ref{PfZ}) survives, namely (using $\Epsilon_1 \cdot \Epsilon_n = -1$)
\be
\Pf {\Psi_{1,n}}  (\bigset)
~=~  (-1)^n \psi_{n+1,2n} \Pf {\Psi}_{1,n,n+1,2n}
~=~ {  (-1)^{n+1} \over \sigma_{1,n} } \Pf {\Psi}_{1,n,n+1,2n} (\smallset)
\ee
and therefore \eqn{EPsiK}  becomes
\be
E(\bigset) ~=~ {1 \over \sigma_{1,n}^2 }\Pf {\Psi}_{1,n,n+1,2n} (\smallset)
\label{Escalar}
\ee
where ${\Psi}_{1,n,n+1,2n}$ denotes the matrix obtained from $\Psi$ by
removing the 1st, $n$th, $(n+1)$st and $(2n)$th rows and columns.
\para

Using \eqn{Escalar} in \eqn{sumCE},
we obtain the $d$-dimensional amplitude for 
two massive adjoint scalars and $n-2$ gauge bosons,
one of which may be massive:
\be
\cA 
~=~   (-1)^{n-1} 
\sumset
{ C(\sigmaset) \over \det'\Phi (\smallsubset)   }  
{ \Pf {\Psi}_{1,n,n+1,2n} (\smallset) \over   \sigma_{1,n}^2   } \,.
\label{massivescalar}
\ee
For the same reasons as before, 
this amplitude is independent of the masses $m_1$, $m_2$, and $m_n$
when written in terms of the invariants (\ref{invariants}).
\para

\medskip\noindent{{\bf Amplitudes for two fundamental scalars 
and $n-2$ gauge bosons}}\smallskip

The scalars in the amplitude (\ref{massivescalar}) transform in the adjoint representation.
The amplitude for scalars in the fundamental representation is essentially the same
except that we replace 
${\bf c}_{1 \gamma n} $ 
in \eqn{defC} with  \cite{Naculich:2014naa}
\be
{\bf t}_{1 \gamma n}  ~=~ 
\left( {T}^{\textsf{a}_{\gamma(2)}}{T}^{\textsf{a}_{\gamma(3)}}
\cdots {T}^{\textsf{a}_{\gamma(n-1)}} \right)^{\textsf{i}_1}_{~~ \textsf{i}_n} 
\ee
where $T^\textsf{a}$ denote generators in the fundamental representation.
In this way, we recover the results of ref.~\cite{Naculich:2014naa} 
for amplitudes with two massive scalars in the fundamental representation
and $n-2$ massless gauge bosons.\footnote{The matrix $\Psi_{1,n,n+1,2n}$ 
is denoted as $\Psi$ in ref.~\cite{Naculich:2014naa}
(note the clarification in {\tt v2}).}
\para

Again, we illustrate our result for the four-point amplitude 
with two fundamental scalars and two gauge bosons.  
On the single solution of the $n=4$ scattering equations, we obtain
\be
 \Pf {\Psi}_{1,4,5,8} (\smallset) 
~\longrightarrow~ {   \kk{3}{4}\over ~\kk{1}{3} ~\kk{2}{3}}  \LL
\label{Efourscalar}
\ee
where 
\be
\LL =
(\kk{1}{ 3} ~ \kk{3}{4}) ~\epseps{2}{3}  
+
\kk{1}{ 3} ~\epsk{2}{1}~ \epsk{3}{4} +\kk{3}{4} ~\epsk{2}{4} ~\epsk{3}{1}  \,.
\label{defL}
\ee
\para
Putting this together with \eqns{Phifour}{Cfour}, we obtain the amplitude
\be 
\cA_{\psi g g \bar\psi}  ~=~ { 1 \over 2 \kk{2}{3} } 
\left[ { {\bf t}_{1234} \over \kk{3}{4} } +  { {\bf t}_{1324} \over \kk{1}{3} } \right] \LL, 
\qquad\qquad \hbox{ when } m_3=0 \,.
\ee
Again we see the expected factorization into a product of a color
factor and a polarization-dependent factor \cite{Zhu:1980sz,Goebel:1980es},
valid provided one of the gauge bosons is massless.
\para

Consider an amplitude involving a photon, $W$, and a pair of scalar quarks
$\psi_d$ and $\psi_{\bar u}$.
In this case, each color factor 
${\bf t}_{1\cdots  4}$
is proportional
to the $\psi_{d} W \psi_{\bar u}$ coupling times 
the charge of the quark interacting with the photon, giving 
\be 
\cA_{\psi_d W \gamma \psi_{\bar u} }  ~\propto~ { 1 \over \kk{2}{3} } 
\left[ { Q_u \over \kk{3}{4} } +  { Q_d  \over \kk{1}{3} } \right] \LL
\ee
agreeing with a Feynman diagram evaluation \cite{Grose:1980cc}.
Note that this amplitude vanishes when 
\be
{ \kk{1}{3} \over \kk{3}{4} } ~=~ - { Q_d \over Q_u }  = \half
\ee
independent of the polarizations.
This implies that the amplitude for 
$ \psi_d \psi_{\bar{u}} \to W^- \gamma  $,
vanishes when $\cos \theta = -1/3$ (for massless quarks),
where $\theta$ is the angle between $\psi_d$ and $W^-$
in the center-of-mass frame.
Such ``radiation zeros'' were first found in calculations of 
$ d \bar{u} \to W^- \gamma  $ processes \cite{Mikaelian:1977ux,Brown:1979ux,Mikaelian:1979nr},
leading to the discovery of four-point factorization \cite{Zhu:1980sz,Goebel:1980es},
which was the forerunner of the BCJ relations \cite{Bern:2008qj}.
Radiation zeros were also found to be present in higher-point
amplitudes under certain conditions \cite{Brodsky:1982sh,Brown:1982xx}.

\section{Gravity amplitudes}
\setcounter{equation}{0}
\label{sec:gravity}

In this section, 
we demonstrate that various gravitational amplitudes involving 
up to three massive particles can be expressed
as a sum over the $(n-3)!$ solutions of the scattering equations,
and that the expressions for these amplitudes, 
when written in terms of the kinematic invariants (\ref{invariants}),
are independent of the masses of the particles. 
\para

We begin in $d+M$ dimensions with the CHY representation
for the tree-level gravitational scattering amplitude 
\cite{Cachazo:2013hca,Cachazo:2013iea}
\be
\cAgv 
~=~ 
(-1)^{n-1} 
\sumset
\frac{ E (\bigset) ~E(\tbigset) }{  \det'\Phi (\bigsubset)  }
\label{sumEE}
\ee
for particles with momenta $K_a$ and whose polarization state is described by a tensor  
$ \Epsilon_a^{\mu\nu} = \Epsilon_a^\mu {\tilde \Epsilon}_a^\nu$,
which includes the graviton, B-field, and dilaton.
The factors 
$\det'\Phi (\bigsubset)$
and 
$ E (\bigset)$ 
were defined in \eqns{detPhi}{EPsiK},
and the sum is over the solutions of the scattering equations (\ref{Kscatt}).
\para

In ref.~\cite{Cachazo:2014xea}, dimensional reduction from 
$d+M$ to $d$ dimensions was used to obtain mixed amplitudes 
in Einstein-Maxwell theory with gauge group $U(1)^M$.
We will use the same approach to obtain various mixed amplitudes involving massive particles.
\para

\medskip\noindent{{\bf Amplitudes for two gauge bosons and $n-2$ gravitons}}\smallskip

To obtain amplitudes for two (possibly massive) gauge bosons 
and $n-2$ gravitons,
we set $K_a = (k_a | \kappa_a)$, with $\kappa_a = 0$ for 
the gravitons ($a=2, \cdots, n-1$). 
We choose the $(d+M)$-dimensional polarization vectors to be 
$\Epsilon_1 = \Epsilon_n = (0 | 1,0,\cdots )$
and 
$\Epsilon_a = (\epsilon_a | 0, 0, \cdots)$ for $a=2, \cdots, n-1$,
so that 
$E(\bigset)$ is given by \eqn{Escalar},
while  we set
${\tilde \Epsilon}_a = ({\tilde\epsilon}_a | 0, 0, \cdots)$ for all $a$
so that $E(\tbigset)$ is given by \eqn{EPsik}.
This yields the amplitude in $d$ dimensions
\be
\cA
~=~
\sumset
{ \Pf {\Psi}_{1,n} (\smallset) ~ \Pf {\Psi}_{1,n,n+1,2n} (\tsmallset)
\over 
\det'\Phi (\smallsubset)   ~\sigma_{1,n}^3   } \,.
\label{massivegaugegravity}
\ee
Although dimensional reduction 
gives the amplitude for two photons and $n-2$ gravitons, 
it remains valid for any gauge theory since no non-abelian vertices are involved.
For the same reasons described previously, 
this amplitude has no dependence 
(when expressed in terms of the invariants (\ref{invariants}))
on the mass of the gauge bosons.
\para

We illustrate \eqn{massivegaugegravity} for the case of 
the four-point amplitude $gh \to gh$,
describing the scattering of gravitons ($h$) from gauge bosons ($g$).
Assembling eqs.~(\ref{Phifour}),  (\ref{Efourgauge}), and (\ref{Efourscalar}),
we obtain 
\be
\cA_{ghhg} 
 ~=~ ~-~ { 1 \over 2 \kk{1}{3}~ \kk{2}{3} ~ \kk{3}{4}}  \LL ~{\tilde \KK}
\label{ghhg}
\ee
where $\LL$ was defined in \eqn{defL}
and $\tilde \KK$ indicates that $\epsilon$ is replaced by $\tilde\epsilon$ in
\eqn{defK}. 
\Eqn{ghhg} agrees with the result of a Feynman diagram calculation \cite{Choi:1994ax}
in which the factorization into $\LL$ and $\tilde\KK$ was noted.
As emphasized in ref.~\cite{Holstein:2006ry},
factorization relates graviton scattering from $W$'s to
Compton scattering from $W$'s, but only when the $W$ magnetic moment
is given by its standard model value $g=2$.

\medskip\noindent{{\bf Amplitudes for two scalars and $n-2$ gravitons}}\smallskip

To obtain amplitudes for two massive scalars and $n-2$ gravitons,
we choose $K_a$ and $\Epsilon_a$ to be the same as above,
but now set
$\tEpsilon_1 = \tEpsilon_n = (0 | 1,0,\cdots )$
and 
$\tEpsilon_a = (\tepsilon_a | 0, 0, \cdots)$ for $a=2, \cdots, n-1$.
Hence both both $E(\bigset)$ and $E(\tbigset)$ are given by \eqn{Escalar},
and the $d$-dimensional amplitude becomes
\be
\cA
~=~  (-1)^{n-1}
\sumset
{ \Pf {\Psi}_{1,n,n+1,2n} (\smallset) ~ \Pf {\Psi}_{1,n,n+1,2n} (\tsmallset)
\over 
\det'\Phi (\smallsubset)   ~\sigma_{1,n}^4   }
\label{massivescalargravity}
\ee
thus recovering the expression obtained in ref.~\cite{Naculich:2014naa}.
As we have seen before, the amplitude is independent of the scalar mass,
when expressed in terms of the invariants (\ref{invariants}).
\para

We illustrate \eqn{massivescalargravity} for the case of 
the scattering $\psi h \to \psi h$ 
of gravitons from massive scalars
\cite{Gross:1968in}. 
Assembling \eqns{Phifour}{Efourscalar}, we obtain
\be
\cA_{\psi h h \bar\psi} ~=~
 ~-~ { 1 \over 2 \kk{1}{3}~ \kk{2}{3} ~ \kk{3}{4}}  \LL ~{\tilde \LL}
\ee
where $\tilde \LL$ indicates that $\epsilon$ is replaced by $\tilde\epsilon$ in
\eqn{defL}. 
The factorization of this amplitude was first noted by the
authors of ref.~\cite{Choi:1994ax}.
\para

\medskip\noindent{{\bf Amplitudes for three gauge bosons 
and $n-3$ gravitons}}\smallskip

Since the amplitude for three (possibly) massive gauge bosons and $n-3$ gravitons
cannot be obtained directly from the dimensional reduction of \eqn{sumEE}
to an Einstein-Maxwell theory (since the triple photon vertex vanishes),
we start instead with the mixed amplitude for three massless gauge bosons and $n-3$ gravitons
in $d+M$ dimensions \cite{Cachazo:2014nsa}
\be
\cA
~=~ (-1)^{n-1} 
\sumset
\frac{ C({\sigma_1, \sigma_2, \sigma_n}) ~E (\Epsilon_3, \cdots, \Epsilon_{n-1}, \bigsubset) ~E(\tbigset) }{  \det'\Phi (\bigsubset)  }
\label{threemassless}
\ee
where
\be
C({\sigma_1, \sigma_2, \sigma_n})
~=~ { \Tr ( T^{\textsf{a}_{1}}T^{\textsf{a}_{2}}  T^{\textsf{a}_{n}} )
\over \sigma_{12} \sigma_{2n} \sigma_{n1} }, \quad
E (\Epsilon_3, \cdots, \Epsilon_{n-1}, \bigsubset) 
 ~=~ 
\Pf \Psi_{1,2,n,n+1,n+2,2n} (\bigset) 
\ee
and
$ \Psi_{1,2,n,n+1,n+2,2n}$
denotes $\Psi$ with the six rows and columns
indicated removed.
\Eqn{threemassless} can be obtained from a pure graviton amplitude 
using a ``squeezing'' process \cite{Cachazo:2014xea}.
\para

To obtain the amplitude for three {\it massive} gauge bosons and $n-3$ gravitons in $d$ dimensions, we 
dimensionally reduce \eqn{threemassless} by taking 
$\Epsilon_a = (\epsilon_a |  0, 0, \cdots)$, 
$\tEpsilon_a = (\tepsilon_a |  0, 0, \cdots)$,
and $K_a = (k_a | \kappa_a)$ for all $a$,
with $\kappa_a=0$ for $a=3, \cdots, n-1$.
Since $K_a \cdot K_b  = k_a \cdot k_b$,
$K_a \cdot \Epsilon_b = k_a \cdot \epsilon_b$, and
$\Epsilon_a \cdot \Epsilon_b = \epsilon_a \cdot \epsilon_b$
for all entries of the matrices
$ \Psi_{1,2,n,n+1,n+2,2n}$ and $ \Psi_{1,n}$,
\eqn{threemassless} reduces to 
\be
\cA
~=~ 
\sumset
{ 
C({\sigma_1, \sigma_2, \sigma_n}) 
~\Pf \Psi_{1,2,n,n+1,n+2,2n} (\smallset) ~\Pf \Psi_{1,n} (\tsmallset)
\over 
  \det'\Phi (\smallsubset)  ~\sigma_{1,n}  }
\label{threemassivegauge}
\ee
which is independent of the gauge boson masses
(when expressed in terms of invariants (\ref{invariants})).
\para

To illustrate \eqn{threemassivegauge}
for the four-point function,
we evaluate its components on the single solution
of the $n=4$ scattering equations to obtain
\ba
C({\sigma_1, \sigma_2, \sigma_4}) 
&\longrightarrow &
{ \fthree \over \sigma_4^2 } ,  \\
\Pf \Psi_{1,2,4,5,6,8} (\smallset)  
~=~ \sum_{c\neq 3}\frac{ \epsk{3}{c} }{\sigma_{3,c}}
&\longrightarrow &
 { \kk{3}{4} \left( \kk{1}{3} ~\epsk{3}{2} - \kk{2}{3} ~\epsk{3}{1}  \right)
\over  \kk{1}{3} ~\kk{2}{3} }
\nonumber
\ea
together with \eqns{Phifour}{Efourgauge}.
Assembling the pieces and taking $\sigma_4 \to\infty$, we obtain
\be
\cA_{g g h g} ~=~
 { 
 \left( \kk{1}{3} ~\epsk{3}{2} - \kk{2}{3} ~\epsk{3}{1}  \right) 
\over  2 \kk{1}{3} ~\kk{2}{3} ~\kk{3}{4} }
 \fthree~
\tilde{\KK}
\label{gghg}
\ee
where all three gauge bosons can be massive.
This result agrees with Feynman diagram computations 
of graviton photoproduction $W \gamma \to W h$ from massive vector bosons
\cite{Choi:1994ax,Bjerrum-Bohr:2014lea}.
\Eqn{gghg} implies that $ W Z^0 \to W h$ should also factorize.

\medskip\noindent{{\bf Amplitudes for two scalars, one gauge boson, and $n-3$ gravitons}}\smallskip

Finally, to obtain the amplitude for two massless scalars, a (possibly) massive gauge boson,
and $n-3$ gravitons in $d$ dimensions, 
we choose $K_a$ and $\Epsilon_a$ to be the same as in the previous example,
but now set
$\tEpsilon_1 = \tEpsilon_n = (0 | 1,0,\cdots )$
and 
$\tEpsilon_a = (\tepsilon_a | 0, 0, \cdots)$ for $a=2, \cdots, n-1$.
Then \eqn{threemassless} reduces to
\be
\cA
~=~ (-1)^{n-1} 
\sumset
{ 
C({\sigma_1, \sigma_2, \sigma_n}) 
~\Pf \Psi_{1,2,n,n+1,n+2,2n} (\smallset) ~\Pf \Psi_{1,n,n+1,2n} (\tsmallset)
\over 
  \det'\Phi (\smallsubset)  ~\sigma_{1,n}^2  }
\ee
once again independent of the masses
(when expressed in terms of invariants (\ref{invariants})).
For the four-point amplitude, this yields
\be
\cA_{\psi g h \bar{\psi}} ~=~
 { 
 \left( \kk{1}{3} ~\epsk{3}{2} - \kk{2}{3} ~\epsk{3}{1}  \right) 
\over  2 \kk{1}{3} ~\kk{2}{3} ~\kk{3}{4} }
 \fthree~
\tilde{\LL}
\label{psighpis}
\ee
which agrees with Feynman diagram computations of graviton 
photoproduction $\psi \gamma \to \psi h$ from scalars \cite{Choi:1994ax,Holstein:2006bh}. 
\Eqn{psighpis} implies that 
$ \psi Z^0 \to \psi h$ 
and 
$ \psi_d W^+ \to \psi_u h$ 
should also factorize.
\para

\section{Conclusions}
\setcounter{equation}{0}
\label{sec:concl}

In this paper we have derived CHY representations for a wide 
class of tree-level gauge and gravitational amplitudes 
containing up to three massive particles.
These mixed amplitudes involve gravitons, massive and massless gauge bosons,
and adjoint or fundamental massive scalar particles.
All of these amplitudes can be obtained through the dimensional
reduction of massless gauge and gravitational amplitudes in a higher dimension.
Further, we demonstrated that amplitudes containing up to three massive particles,
when written in terms of the kinematic invariants (\ref{invariants}),
are in fact independent of the masses,
and so have the same form as purely massless amplitudes.
\para

It is not so straightforward to obtain CHY representations for 
gauge and gravitational amplitudes with four or more massive particles.
As explained in sec.~\ref{sec:scattering}, such amplitudes do not generically
result from the dimensional reduction of massless amplitudes
because of the difficulty of satisfying momentum conservation in the internal dimensions.
Moreover, the propagator matrix in this case generally has rank
greater than $(n-3)!$ and so cannot be represented as a sum over 
$(n-3)!$ solutions.
\para

On the other hand, there is good reason to suspect the existence of
CHY representations involving fermions,
since the factorization of four-point gauge amplitudes 
also holds for amplitudes containing spin-one-half particles
\cite{Zhu:1980sz,Goebel:1980es}
and gravitational scattering from spin-one-half particles
\cite{Choi:1994ax,Holstein:2006bh,Holstein:2006ry,Bjerrum-Bohr:2014lea}
has also been shown to factorize.
For progress in this area, see refs.~\cite{Adamo:2013tsa,
Bjerrum-Bohr:2014qwa,Weinzierl:2014ava}.
\para

\setcounter{equation}{0}
\section*{Acknowledgments}
This material is based upon work supported by the 
National Science Foundation under Grant No.~PHY14-16123.

\vfil\break

\end{document}